\documentclass[12pt,%preprint,
aps,amsmath,byrevtex,prd,titlepage,
preprintnumbers,
tightenlines,
nofootinbib]{revtex4-1}

\usepackage{amsmath}
\usepackage{amssymb}
\usepackage{bm}
\usepackage{hyperref}
\usepackage{graphicx}
\usepackage{slashed}
\usepackage{dcolumn}

\newcommand{\beq}{\begin{equation}}
\newcommand{\eeq}{\end{equation}}
\newcommand{\eq}[1]{Eq.~(\ref{#1})}

\begin{document}

\title {Three-Loop Contributions to Hyperfine Splitting: Muon Loop Light-by-Light Insertion and Other Closed Lepton Loops}
\author {Michael I. Eides}
\altaffiliation[Also at ]{the Petersburg Nuclear Physics Institute,
Gatchina, St.Petersburg 188300, Russia}
\email[Email address: ]{eides@pa.uky.edu, eides@thd.pnpi.spb.ru}
\affiliation{Department of Physics and Astronomy,
University of Kentucky, Lexington, KY 40506, USA}
\author{Valery A. Shelyuto}
\email[Email address: ]{shelyuto@vniim.ru}
\affiliation{D. I.  Mendeleyev Institute for Metrology,
St.Petersburg 190005, Russia}
%\date{}

\begin{abstract}
The muon and tauon light-by-light scattering contributions to hyperfine splitting in muonium are calculated. These results conclude calculation of all hard three-loop contributions to hyperfine splitting containing graphs with closed fermion loops. We discuss the special role that the lepton anomalous magnetic moments play in these calculations. The full result for all three-loop radiative-recoil corrections to hyperfine splitting generated by the graphs with closed lepton loops is presented.

\end{abstract}

%\pacs{31.30.jf,32.10.Fn,36.10.Ee}
%\keywords{hyperfine splitting}

\preprint{UK/14-06}

\maketitle

\section{Introduction}

Calculation of high order corrections to hyperfine splitting in muonium is a classic playground of high precision bound state quantum electrodynamics. For many years theory and experiment developed hand in hand, and the measurements of the hyperfine splitting (HFS) were the best source for the precise value of the electron-muon mass ratio (see, e.g, reviews \cite{egs2001,egs2007,mtn2012}. The current experimental error of HFS in muonium is in the interval 16-53 Hz ($1.2-3.6\times10^{-8}$) \cite{mbb,lbdd}. More than ten years ago we declared reduction of the theoretical error of HFS in muonium to the level of 10 Hz to be an achievable goal of the theoretical research \cite{egs2001,egs2007}. This goal became recently even more pressing in view of a new high accuracy measurement of muonium HFS planned now at J-PARC, Japan \cite{sasaki,tanaka}. The goal of this experiment is to reduce the experimental error  by an order of magnitude, to the level of a few parts per billion, what is below 10 Hz.

In order to reduce the theoretical error below 10 Hz    one has to calculate single-logarithmic and nonlogarithmic in mass ratio hard radiative-recoil corrections of order $\alpha^2(Z\alpha)(m/M)\widetilde E_F$, as well as soft nonlogarithmic contributions of orders $(Z\alpha)^3(m/M)\widetilde E_F$ and $\alpha(Z\alpha)^2(m/M)\widetilde E_F$\footnote{Here $\alpha$ is the fine structure constant, $m$ and $M$ are the electron and muon masses, respectively. $Z=1$ is the charge of the constituent muon, it is convenient to introduce it for classification of different contributions. The Fermi energy is defined as $\widetilde E_F=(8/3)(Z\alpha)^4(m/M)({m_r}/{m})^3m$, where $m_r=mM/(m+M)$ is the reduced mass.}. We have concentrated our efforts on calculation of hard radiative-recoil corrections of order $\alpha^2(Z\alpha)(m/M)\widetilde E_F$, and in recent years calculated all single-logarithmic and nonlogarithmic correction the HFS arising from the diagrams with closed lepton loops \cite{egs2001rr,egs2003,egs2004,egs2005,es2009,prd2009pr,jetp2010,es2013,es2014,es2014_2}. Below we will present the details of the recent calculation of the last previously unknown light-by-light scattering contribution to HFS arising from the virtual muon and tauon loops\footnote{The results of this calculation were already reported in \cite{es2014_2}.}. We will also discuss radiative-recoil corrections connected with the anomalous magnetic moments and present complete results for all corrections of order $\alpha^2(Z\alpha)(m/M)\widetilde E_F$ generated by the three-loop diagrams containing closed lepton loops.

\section{Muon and Tauon Loop Light-by-Light Insertions}

\subsection{General Expressions and the Infrared Problems}

\begin{figure}[htb]
\includegraphics[height=1.cm]{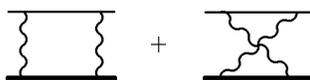}
\caption{\label{twoph}Diagrams with two-photon exchanges}
\end{figure}

Radiative insertions in the diagrams with two-photon exchanges in Fig. \ref{twoph} generate all three-loop diagrams for the contributions of order $\alpha^2(Z\alpha)(m/M)\widetilde E_F$. It is well known that in any gauge invariant set of diagrams radiative insertions suppress integration momenta small in comparison with the electron mass. As a result the characteristic integration momenta in these diagrams are of order of the electron mass or higher, these are hard corrections. This significantly simplifies calculations because then we can neglect momenta of the external wave functions and calculate the diagrams in the scattering approximation with the on-shell external momenta. The contribution to HFS is obtained by projecting the diagrams on the HFS spin structure and multiplying the result by the value of Schr\"odinger-Coulomb wave function at the origin squared (for more details see, e.g. \cite{egs2001,egs2007}).

\begin{figure}[htb]
\includegraphics
[height=2.5cm]
{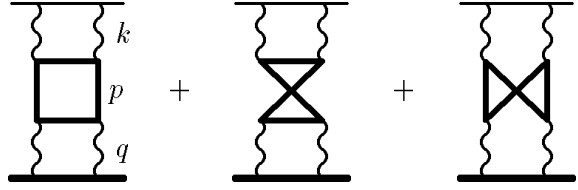}
\caption{\label{lblrec}
Diagrams with the muon (tauon) light-by-light scattering block}
\end{figure}

The general expression for the muon loop light-by-light (LBL)  scattering contribution to HFS  in Fig. \ref{lblrec} is similar to the respective electron loop contribution (see, e.g., \cite{es2013,es2014}), and can be written in the form

\beq
\Delta E=\frac{\alpha^2(Z\alpha)}{\pi^3}\frac{m}{M}\widetilde E_FJ,
\eeq

\noindent
where  $J$ is a dimensionless integral

\beq \label{jint}
J=-\frac{3M^2}{128}\int \frac{d^4k}{i\pi^2 k^4}
\left(\frac{1}{k^2+2mk_0}+\frac{1}{k^2-2mk_0}\right)T(k^2,k_0).
\eeq

\noindent
The dimensionless function $T(k^2,k_0)$ is a sum of the ladder and crossed diagrams contributions in Fig. \ref{lblrec}

\beq \label{dimesfuntlc}
T(k^2,k_0)=2T_L(k^2,k_0)+T_C(k^2,k_0).
\eeq

\noindent
Explicit expressions for the functions $T_L(k^2,k_0)$ and $T_C(k^2,k_0)$ can be obtained  by the substitution $m\to M$, $q_\mu\to k_\mu$  from the respective formulae in \cite{es2013,es2014}, where these functions were calculated in the case of the electron LBL scattering block.

Only the even in $k_0$ terms in the function $T(k^2,k_0)$ contribute to the integral in \eq{jint}. After rescaling of the integration momentum $k\to kM$, the Wick rotation, and  symmetrization of the function $T(k^2,k_0)$ with respect to $k_0$, $T(k^2,k_0)\to T(k^2,k_0^2)$, the integral in \eq{jint} turns into

\beq \label{wickene}
J=\frac{3}{32\pi}\int_0^\infty \frac{dk^2}{k^2}\int_0^\pi d\theta\sin^2\theta\frac{T(k^2,\cos^2\theta)}{k^2+16\mu^2\cos^2\theta},
\eeq

\noindent
where we have parameterized the Euclidean four-vectors as $k_0=k\cos\theta$, $|\bm k|=k\sin\theta$, $\mu=m/(2M)$, and the function  $T(k^2,\cos^2\theta)$ is the same function as in \eq{jint} but symmetrized with respect to $k_0$ and with the Wick rotated momenta. The dimensionless function $T(k^2,\cos^2\theta)$ after rescaling depends on the dimensionless momentum $k$ and does not contain any parameters with the dimension of mass. Below we will often write the integral in \eq{wickene} as a sum

\beq
J=2J_L+J_R,
\eeq

\noindent
where the terms on the RHS correspond to the respective terms on the RHS in \eq{dimesfuntlc}.

We are looking for the $\mu$-independent contributions generated by the integral in \eq{wickene}. The term with $\mu^2$ in the denominator is irrelevant at large $k$, and the integral is convergent at large $k$ due to the ultraviolet convergence of all diagrams with the LBL insertions.  The case of the small integration momenta is more involved. Due to gauge invariance, the LBL block is strongly suppressed at $k\to0$, and we expect that the integral in \eq{wickene} remains finite even at $\mu=0$ zero. As a result of this finiteness the diagrams in Fig. \ref{lblrec} should not generate either nonrecoil or logarithmically enhanced recoil contributions to HFS in accordance with our physical expectations. However,  at $\mu=0$ convergence  of the small integration momenta contributions from individual diagrams cannot be taken for granted, and we have to consider separate entries in more detail. The functions $T_L(k^2,k_0)$ and $T_C(k^2,k_0)$ are sums of terms each of which is a  multidimensional integral over the Feynman parameters and an explicit function of the integration momentum squared $k^2$ and the integration angle $\theta$. The dependence on angles can be easily separated and therefore we can explicitly calculate the integrals over angles. All these integrals are proportional to one of the two standard functions $\bar{\Phi}_0$ and $\bar{\Phi}_1$ (compare analogous functions in the case of the virtual electron light-by-light scattering loop in \cite{es2014}):

\beq  \label{int-cos0}
\begin{split}
\bar{\Phi}_0& =
\frac{2}{\pi}\int_0^{\pi}{d\theta}
\frac{\sin^2{\theta}}{k^2+16\mu^2\cos^2{\theta}} =
\frac{1}{8\mu^2}\biggl[\frac{1}{k}\sqrt{k^2+16\mu^2}-1\biggr],
\\
\bar{\Phi}_1& =
\frac{2}{\pi}\int_0^{\pi}{d\theta}
\frac{\sin^2{\theta}\cos^2{\theta}}{k^2+16\mu^2\cos^2{\theta}}
=\frac{1}{8\mu^2}\biggl[-\frac{k}{16\mu^2}\sqrt{k^2+16\mu^2}
+\frac{k^2}{16\mu^2}+\frac{1}{2}\biggr].
\end{split}
\eeq

\noindent
The ultraviolet asymptotics of these functions coincide with their exact values at $\mu=0$

\beq \label{uvas}
\bar{\Phi}_{0|\mu=0} = \frac{1}{k^2},\qquad
\bar{\Phi}_{1|\mu=0} =\frac{1}{4k^2}.
\eeq

\noindent
Using these integrals and the explicit expressions for the large momentum asymptotic behavior of the functions $T_{L(C)}(k^2,\cos^2\theta$ (see \cite{es2013}) we once again confirm that the momentum integral in \eq{wickene} is ultraviolet finite.

The infrared region requires more attention. The functions $T_L(k^2,\cos^2\theta)$ and $T_C(k^2,\cos^2\theta)$ contain terms that  decrease only as $k^2$ at small momenta. We would obtain logarithmically infrared divergent integrals if we substitute in the momentum integral in \eq{wickene} such terms together with the angular integrals in \eq{uvas}. These are fake divergences since at $\mu\neq0$

\beq
\bar{\Phi}_{0} \approx \frac{1}{2\mu k}-\frac{1}{8\mu^2}+\dots,
\qquad
\bar{\Phi}_{1} \approx \frac{1}{16\mu^2}-\frac{k}{32\mu^3}+\dots,
\eeq

\noindent
and then the momentum integrals of separate terms in \eq{wickene} are infrared finite. The would be logarithmic divergences are cutoff by the parameter $\mu$ (the upper integration limit is irrelevant for the discussion of the infrared convergence)

\beq
\int_0^1\frac{dk^2}{k^2} \bar{\Phi}_{0}k^2\approx 2\ln\frac{1}{2\mu}+1+{\cal O}(\mu),
\qquad
\int_0^1\frac{dk^2}{k^2} \bar{\Phi}_{1}k^2\approx \frac{1}{2}\ln\frac{1}{2\mu}-\frac{1}{8}+{\cal O}(\mu).
\eeq

\noindent
We see that one cannot delete $\mu$ in the integrals of separate terms in \eq{wickene} without generating artificial infrared divergences. On the other hand, we know that due to gauge invariance all infrared logarithms cancel in the total integral in \eq{wickene} that remains finite when $\mu$ goes to zero. We are interested only in the value of the integral at $\mu=0$, so our next goal is to organize the calculations in such way that allows to let $\mu=0$ before integration. This approach leads to significant simplification of numerical calculations since the integrals with $\mu\neq0$ are much more involved and the final result arises as a result of cancelation of big numbers.

To facilitate further calculations we represent the functions $T_{L(C)}(k^2,\cos^2\theta)$ in the form

\beq
T_{L(C)}(k^2,\cos^2\theta)=T^{reg}_{L(C)}(k^2,\cos^2\theta)
+T^{sing}_{L(C)}(k^2,\cos^2\theta),
\eeq

\noindent
where the functions $T^{reg}$'s decrease faster than $k^2$  at small $k^2$, and the functions $T^{sing}$'s decrease as $k^2$  at small $k^2$.

In these terms the integral in \eq{wickene}  has the form

\beq
J=J^{reg}+J^{sing},
\eeq

\noindent
where

\beq \label{singregint}
J^{reg(sing)}=\frac{3}{32\pi}\int_0^\infty \frac{dk^2}{k^2}\int_0^\pi d\theta\sin^2\theta\frac{T^{reg(sing)}(k^2,\cos^2\theta)}{k^2+16\mu^2\cos^2\theta},
\eeq

\noindent
and

\beq
T^{reg(sing)}(k^2,\cos^2\theta)=2T^{reg(sing)}_{L}(k^2,\cos^2\theta)
+T^{reg(sing)}_{C}(k^2,\cos^2\theta).
\eeq

\subsection{Calculation of the Infrared Safe Integrals}

Consider first calculation of the infrared safe integrals in \eq{singregint}. One can obtain an explicit expression for the infrared safe function $T^{reg}$ by omitting all terms in the explicit representations for the functions $T_{L(C)}$ in \cite{es2013,es2014} that decrease as $k^2$ or $k_0^2$ at small $k$. To preserve the formulae relatively compact we put down explicit expressions for the functions $T^{reg}$ as they are used in \eq{dimesfuntlc}, before the Wick rotation and symmetrization. Then the regular ladder function has the form of a sum of nine multidimensional integrals

\beq  \label{genint-L}
T^{reg}_{L}(k^2, k_0)=\frac{128}{3}\int_0^1 {dy} \int_0^1 {dz}\int_0^1 {du}\int_0^1 {dt}
\sum_i {\cal T}^{reg}_{L,i}(y, z, u, t, k^2, k_0),
\eeq

\noindent
where

\beq \label{calTL-1}
\begin{split}
{\cal T}^{reg}_{L,1}& =  yz(1-t)(1-u)^2
\Biggr\{-\frac{(2k^2+k_0^2)k^2 d^2}{\Delta^2}
- \frac{ k_0  (5k^2+k_0^2) \tau d}{\Delta^2} \Biggr\},
%\end{split}
%\eeq
%\beq \label{calTL-2}
\\
{\cal T}^{reg}_{L,2}&
=-\frac32(2k^2+k_0^2) \frac{y^2z^2(1-z) k^2}{(1-y)^2}
\frac{(1-t) u(1-u)^2}{\Delta^2},
%\eeq
%\beq \label{calTL-3}
%\begin{split}
\\
{\cal T}^{reg}_{L,3}& =\Biggl\{\frac{(1-2y) +2yz}{1-y} \frac{(1-t)(1-u)^2}{\Delta^2}
-(1-z) \frac{u(1-u)}{\Delta^2}
\\
&+2 \frac{y^2z^2(1-z) k^2}{(1-y)^2}
\frac{(1-t) u(1-u)^2}{\Delta^3}\Biggr\} (2k^2+k_0^2) k^2d^2,
%\end{split}
%\eeq
%\beq \label{calTL-4}
%\begin{split}
\\
{\cal T}^{reg}_{L,4}& =
2\frac{y^2z^2(1-z) k^2}{(1-y)^2}
\frac{(1-t) u(1-u)^2}{\Delta^3}(2k^2+k_0^2)  \tau^2
\\
&
+\Biggl\{\frac{(1-2y) +2yz}{1-y} \frac{(1-t)(1-u)^2}{\Delta^2}
-(1-z) \frac{u(1-u)}{\Delta^2}
\\
&
+2 \frac{y^2z^2(1-z) k^2}{(1-y)^2}
\frac{(1-t) u(1-u)^2}{\Delta^3}\Biggr\}
k_0 (5k^2+k_0^2) \tau  d,
%\end{split}
%\eeq
%\beq \label{calTL-5}
\\
{\cal T}^{reg}_{L,5}&=0,
%\eeq
%\beq \label{calTL-6}
%\begin{split}
\\
{\cal T}^{reg}_{L,6}&  =4\int_0^1 {d\xi}\xi  yz^2 (1-t) u(1-u)^2
\\
&\times\Biggl\{\biggl[\frac34 \frac{1}{\Delta_{\xi}^2} - \frac{k^2d^2_{\xi}}{\Delta_{\xi}^3}\biggr]
(2k^2+k_0^2) k^2 d_{\xi}
- \frac{\tau^2  k_0^2 d_{\xi}}{\Delta_{\xi}^3}  (8k^2+k_0^2)
\\
&+\biggl[\frac14 \frac{1}{\Delta_{\xi}^2} - \frac{k^2d^2_{\xi}}{\Delta_{\xi}^3}\biggr]
(7k^2+2k_0^2) k_0\tau - \frac{3 k_0^3 \tau^3}{\Delta_{\xi}^3}\Biggr\},
%\end{split}
%\eeq
%\beq \label{calTL-7}
%\begin{split}
\\
{\cal T}^{reg}_{L,7}& =-\frac{yz(1-z)}{1-y}\frac{k^2 u(1-u)}{\Delta^2}
\Bigl[(2k^2+k_0^2) d + 3 k_0\tau \Bigr],
%\end{split}
%\eeq
%\beq \label{calTL-8}
%\begin{split}
\\
{\cal T}^{reg}_{L,8}& = 2 \frac{yz(1-z)}{1-y} (1-t) u(1-u)^2
\Biggl\{\biggl[-\frac34 \frac{1}{\Delta^2} + \frac{k^2d^2}{\Delta^3}\biggr]
(2k^2+k_0^2) k^2 d
\\
&+ \frac{\tau^2  k_0^2 d}{\Delta^3}  (8k^2+k_0^2)
+\biggl[-\frac14 \frac{1}{\Delta^2} + \frac{k^2d^2}{\Delta^3}\biggr]
(7k^2+2k_0^2) k_0\tau + \frac{3 k_0^3 \tau^3}{\Delta^3}\Biggr\},
%\end{split}
%\eeq
%\beq
%\label{calTL-9}
%\begin{split}
\\
{\cal T}^{reg}_{L,9}&  =4 \frac{yz(1-z)}{1-y}(1-t) u(1-u)^2
\biggl[-\frac{1}{4} \frac{1}{\Delta^2} (2k^2+k_0^2) k^2d
\\
&+k^2(k^2-k_0^2) \frac{\tau^2 d}{\Delta^3}
- \frac{1}{4} \frac{1}{\Delta^2}(2k^2+k_0^2) k_0 \tau
+k_0(k^2-k_0^2) \frac{\tau^3}{\Delta^3}\biggr].
\end{split}
\eeq

\noindent
The regular  crossed diagram contribution reduces to one-multidimensional integral

\beq  \label{genint-C}
T^{reg}_{C}(k^2, k_0)=\frac{128}{3}\int_0^1 {dx}\int_0^1 {dy} \int_0^1 {dz}\int_0^1 {du}\int_0^1 {dt}{\cal T}^{reg}_{C}(x, y, z, u, t, k^2, k_0),
\eeq

\noindent
where (${\cal T}^{reg}_{C,2}=0$)

\beq \label{calTC}
\begin{split}
{\cal T}^{reg}_{C}&={\cal T}^{reg}_{C,1}+{\cal T}^{reg}_{C,3}
\\
& =\frac12 \frac{x(1-t)(1-u)^2}{1-xy}
\Biggr[-4(k^2-k_0^2) \frac{ k^2 \tau^2  d^2}{\Delta^3}
-3k^2\frac{k_0 \tau d}{\Delta^2}
-4(k^2-k_0^2) \frac{k_0 \tau^3  d}{\Delta^3}\Biggr].
\end{split}
\eeq

\noindent
In \eq{calTL-1}-\eq{calTC}

\beq  \label{Delta}
\begin{split}
&\Delta = g\Bigl[-k^2 + 2bk_0 + a^2\Bigr],
\quad
a^2=\frac{1}{g}\biggl[\tau^2+\frac{M^2u}{xy(1-xy)}\biggr],\quad
b=\frac{\tau d}{g},
\\
&d=\xi u\biggl[z- \frac{1-x}{1-xy}\biggr],\qquad \tau =M(1-u)t,
\qquad
g=g_0-d^2,
\\
&g_0=\frac{u(1-yz)(1-x+xyz)}{y(1-xy)},
\end{split}
\eeq

\noindent
and $x=1$ in \eq{calTL-1},
%-\eq{calTL-9}
while $\xi=1$ in all functions in \eq{calTL-1} and \eq{calTC} except
%\eq{calTL-6}
the function ${\cal T}^{reg}_{L,4}$.

We still need to make the Wick rotation and symmetrize the functions $T^{reg}$ over $k_0$. Details of similar symmetrization are described in \cite{es2014}, and we will not describe them here. After symmetrization we substitute the functions $T^{reg}$ in the integral $J^{reg}$ in \eq{singregint}. We can safely let $\mu=0$ before integration, what makes calculation straightforward. We collected the separate contributions in Table \ref{regint}. Summing all these regular contribution we obtain

\beq \label{jreg}
J^{reg}=2\sum_{i=1}^9J^{reg}_{L,i}+J^{reg}_C=-2.146\,39\,(3).
\eeq

\begin{table}[htb]
\caption{\label{regint}Regular Integrals}
\begin{ruledtabular}
\begin{tabular}{|l|d|}
$J^{reg}_{L,1}$ &
-\,0.\,014\,805\,(4)
\\
$J^{reg}_{L,2}$ & -\,1.\,202\,396\,(4)
\\
$J^{reg}_{L,3}$ & 0.\,050\,539\,(1)
\\
$J^{reg}_{L,4}$ &
0.\,162\,208\,(6)
\\
$J^{reg}_{L,6}$ &
0.\,014\,963\,(4)
\\
$J^{reg}_{L,7}$ & -0.\,045\,490\,(3)
\\
$J^{reg}_{L,8}$ &
-\,0.\,023\,826\,(6)
\\
$J^{reg}_{L,9}$ & -\,0.\,014\,441\,(4)
\\
$J^{reg}_{C}$ & 0.\,000\,106\,(1)
\end{tabular}
\end{ruledtabular}
\end{table}

\subsection{Calculation of the Apparently Infrared Singular Integrals}

Consider now the apparently infrared singular integrals $J^{sing}$.  The respective functions $T^{sing}_{L(C)}$ are again sums of multidimensional integrals similar to the ones in \eq{genint-L} and \eq{genint-C}. The functions   ${\cal T}^{sing}_{L(C),i}$ contain those terms from the general functions  ${\cal T}_{L(C),i}$ (see \cite{es2013,es2014}) that were not included in the regular functions in \eq{calTL-1}
%-\eq{calTL-9}
and \eq{calTC}. Explicitly

\beq \label{calTLs-1}
\begin{split}
{\cal T}^{sing}_{L,1}& =  yz(1-t)(1-u)^2
\Biggr\{\frac{2k^2+k_0^2}{\Delta}
- \frac{ (k^2+2k_0^2)\tau^2}{\Delta^2}\Biggr\},
%\end{split}
%\eeq
%\beq \label{calTLs-2}
\\
{\cal T}^{sing}_{L,2}& =\frac32 (2k^2+k_0^2)\Biggl\{-\frac{(1-2y) +2yz}{1-y}
\frac{(1-t)(1-u)^2}{\Delta}+(1-z) \frac{u(1-u)}{\Delta}
\Biggr\},
%\eeq
%\beq \label{calTLs-4}
\\
{\cal T}^{sing}_{L,4}& =(2k^2+k_0^2)  \tau^2
\Biggl\{\frac{(1-2y) +2yz}{1-y} \frac{(1-t)(1-u)^2}{\Delta^2}
-(1-z) \frac{u(1-u)}{\Delta^2}
\Biggr\},
%\eeq
%\beq \label{calTLs-5}
%\begin{split}
\\
{\cal T}^{sing}_{L,5}& =\frac{M^2}{1-y}\frac{(1-t)(1-u)^2}{\Delta^2}
\Bigl[(2k^2+k_0^2) d + 3 k_0\tau \Bigr],
%\end{split}
%\eeq
%\beq \label{calTLs-36789}
\\
{\cal T}^{sing}_{L,3}& =
{\cal T}^{sing}_{L,6}=
{\cal T}^{sing}_{L,7} =
{\cal T}^{sing}_{L,8}=
{\cal T}^{sing}_{L,9}  =0,
%\eeq
%\beq \label{calTCs-1}
\\
{\cal T}^{sing}_{C,1}& =\frac{x(1-t)(1-u)^2}{1-xy}
\Biggr[\frac{2k^2+k_0^2}{\Delta}
- \frac{3}{2}\frac{ k^2\tau^2}{\Delta^2}\Biggr],
%\eeq
%\beq  \label{calTCs-2}
\\
{\cal T}^{sing}_{C,2}& = \frac{x(1-t)(1-u)^2}{1-xy}
\frac{u M^2}{xy(1-xy)}
\biggl[\frac{2k^2+k_0^2}{\Delta^2}
-4 \frac{(k^2-k_0^2) \tau^2}{\Delta^3} \biggr],
%\eeq
%\beq
%\label{calTCs-3}
\\
{\cal T}^{sing}_{C,3}& =0.
\end{split}
\eeq

The integrals of these functions  in \eq{genint-L} and \eq{genint-C}
generate the terms proportional to  $k^2$ and also the terms with higher powers of $k^2$. We would like to separate the terms proportional to $k^2$ (only such terms generate infrared logarithms after momentum integration) and check that all such terms cancel before integration over  $k$. This separation can be achieved with the help of the substitution $\Delta=g(k^2+2bk_0+a^2)\to ga^2\equiv \widetilde a^2$. After this substitution the integrals over the Feynman parameters are quadratic in $k$. Let us demonstrate this explicitly. After the substitution  $\Delta\to \widetilde a^2$ the nonvanishing functions in \eq{calTLs-1}
%-\eq{calTCs-3}
acquire the form

\beq \label{calTLlog-1}
\begin{split}
{\cal T}^{quadr}_{L,1}& = (2k^2+k_0^2)A_{11}+(k^2+2k_0^2)A_{13},
%\eeq
%\beq \label{calTLlog-2}
\\
{\cal T}^{quadr}_{L,2}& =(2k^2+k_0^2)(A_{21}+A_{22}),
%\eeq
%\beq \label{calTLlog-4}
\\
{\cal T}^{quadr}_{L,4}& =(2k^2+k_0^2)(A_{41}+A_{42}),
%\eeq
%\beq \label{calTLlog-5}
\\
{\cal T}^{quadr}_{L,5}& =(2k^2+k_0^2)A_{51} + k^2_0A_{52},
%\eeq
%\beq \label{calTClog-1}
\\
{\cal T}^{quadr}_{C,1}& =(2k^2+k_0^2)B_{11}+k^2B_{13},
%\eeq
%\beq  \label{calTClog-2}
\\
{\cal T}^{quadr}_{C,2}& = (2k^2+k_0^2)B_{21}+(k^2-k_0^2)B_{22},
\end{split}
\eeq

\noindent
where

\beq  \label{calTL-11}
\begin{split}
&A_{11} =
yz(1-t)(1-u)^2\frac{1}{\widetilde{a}^2},
%\eeq
%\beq  \label{calTL-13}
\\
&A_{13} = -
yz(1-t)(1-u)^2\frac{\tau^2}{\widetilde{a}^4}
%\eeq
%\beq  \label{calTL-21-22}
\\
&A_{21}+A_{22}
=\frac32\Biggl[-\frac{(1-2y) +2yz}{1-y}
\frac{(1-t)(1-u)^2}{\widetilde{a}^2}+(1-z)
\frac{u(1-u)}{\widetilde{a}^2}\Biggr],
%\eeq
%\beq  \label{calTL-41-42}
\\
&A_{41}+A_{42} =\tau^2\Biggl[\frac{(1-2y)
+2\,yz}{1-y}  \frac{(1-t)(1-u)^2}{\widetilde{a}^4}-(1-z)
\frac{u(1-u)}{\widetilde{a}^4}\Biggr],
%\eeq
%\beq  \label{calTL-51}
\\
&A_{51}
=\frac{d}{1-y}\frac{(1-t)(1-u)^2}{\widetilde{a}^4},
%\eeq
%\beq  \label{calTL-52}
\\
&A_{52}
=-12\frac{d\tau^2}{1-y}\frac{(1-t)(1-u)^2}{\widetilde{a}^6},
%\eeq
%\beq  \label{calTC-11}
\\
&B_{11}
=\frac{x(1-t)(1-u)^2}{1-xy}\frac{1}{\widetilde{a}^2},
%\eeq
%\beq  \label{calTC-13}
\\
&B_{13}
=-\frac32\frac{x(1-t)(1-u)^2}{1-xy}\frac{\tau^2}{\widetilde{a}^4},
%\eeq
%\beq  \label{calTC-21}
\\
&B_{21} = \frac{\,(1-t)(1-u)^2u}{y(1-xy)^2}
\frac{1}{\widetilde{a}^4},
%\eeq
%\beq  \label{calTC-22}
\\
&B_{22} =
-4\frac{\,(1-t)(1-u)^2u}{y(1-xy)^2}
\frac{\tau^2}{\widetilde{a}^6}.
\end{split}
\eeq

We have calculated the integrals $T^{quadr}_{L(C),ij}$ of the functions in \eq{calTLlog-1} analytically (they are defined similarly to the integrals in \eq{genint-L} and \eq{genint-C}), and the results are collected in Table \ref{loginmt}, where the value of an integral over the Feynman parameters corresponds to the respective coefficient function $A_{ij}$ or $B_{ij}$.

\begin{table}[thhp!]
\caption{\label{loginmt} Integrals for Coefficients before $k^2$}
\begin{ruledtabular}
\begin{tabular}{|l|l||l|l|}
$A_{11}$ &$T^{quadr}_{L,11}=-\frac{\pi^2}{72}+\frac{5}{24}$     &
$B_{11}$ &$T^{quadr}_{C,11}=-\frac{\pi^2}{18}+\frac{5}{6}$
\\
$A_{13}$ &$T^{quadr}_{L,13}=\frac{\pi^2}{36}-\frac{7}{24}$      &
$B_{13}$ &$T^{quadr}_{C,13}=\frac{\pi^2}{6}-\frac{7}{4}$
\\
$A_{21}$ &$T^{quadr}_{L,21}=\frac{\pi^2}{12}-\frac{5}{4}$       &
$B_{21}$ &$T^{quadr}_{C,21}=\frac{\pi^2}{18}-\frac{1}{3}$
\\
$A_{22}$ &$T^{quadr}_{L,22}=-\frac{\pi^2}{12}+\frac{7}{8}$      &
$B_{22}$ &$T^{quadr}_{C,22}=-\frac{7\pi^2}{72}+\frac{5}{6}$
\\
$A_{41}$ &$T^{quadr}_{L,41}=-\frac{\pi^2}{9}+\frac{7}{6}$       &&
\\
$A_{42}$ &$T^{quadr}_{L,42}=\frac{23\pi^2}{288}-\frac{19}{24}$  &&
\\
$A_{51}$ &$T^{quadr}_{L,51}=\frac{\pi^2}{72}-\frac{1}{12}$      &&
\\
$A_{52}$ &$T^{quadr}_{L,52}=-\frac{7\pi^2}{96}+\frac{5}{8}$     &&
\end{tabular}
\end{ruledtabular}
\end{table}

Collecting the results in Table \ref{loginmt} we confirm that the total coefficient before the $k^2$  in the small $k$ expansion of the integrand in \eq{singregint} for $J^{sing}$ is equal zero

\beq \label{canclel}
T^{quadr}=2T^{quadr}_L+T^{quadr}_C=0.
\eeq

\noindent
Let us use this observation to get rid of $\mu$ in the integral for $J^{sing}$ in \eq{singregint} before integration. According to \eq{dimesfuntlc}

\beq
T^{sing}(k^2,\cos^2\theta)=2T^{sing}_L(k^2,\cos^2\theta)+T^{sing}_C(k^2,\cos^2\theta).
\eeq

\noindent
Using the cancelation in \eq{canclel} we observe

\beq
\begin{split}
&T^{sing}(k^2,\cos^2\theta)=
T^{sing}(k^2,\cos^2\theta)-T^{quadr}(k^2,\cos^2\theta)
%\eeq
%\[
\\
&=2\left(T^{sing}_L(k^2,\cos^2\theta)-T^{quadr}_L(k^2,\cos^2\theta)\right)+\left(T^{sing}_C(k^2,\cos^2\theta)
-T^{quadr}_C(k^2,\cos^2\theta)\right)
%\]
%\[
\\
&=2\sum_{ij}\left(T^{sing}_{L,ij}(k^2,\cos^2\theta)-T^{quadr}_{L,ij}(k^2,\cos^2\theta)\right)
+\sum_{ij}\left(T^{sing}_{C,ij}(k^2,\cos^2\theta)
-T^{quadr}_{C,ij}(k^2,\cos^2\theta)\right).
%\]
\end{split}
\eeq

\noindent
Each expression in the brackets on the RHS decreases at small $k$ faster than $k^2$ and therefore allows us to safely let $\mu=0$ in the integral for $J^{sing}$ before integration

\beq
\begin{split}
J^{sing}&=\frac{3}{32\pi}\int_0^\infty {dk^2}\int_0^\pi d\theta\sin^2\theta
\biggl[
2\sum_{ij}\frac{T^{sing}_{L,ij}(k^2,\cos^2\theta)-T^{quadr}_{L,ij}(k^2,\cos^2\theta)}{k^4}
%\eeq
%\[
\\
&+\sum_{ij}\frac{T^{sing}_{C,ij}(k^2,\cos^2\theta)
-T^{quadr}_{C,ij}(k^2,\cos^2\theta)}
{k^4}\biggr].
%\]
\end{split}
\eeq

The problems with the infrared convergence arise only at $k\to0$. We can further simplify the calculations by arbitrarily separating the integration regions of small and large momenta and omitting the term  $T^{quadr}(k^2,\cos^2\theta)=0$ in the large integration momenta region

\beq \label{singiniots}
\begin{split}
J^{sing}&=\frac{3}{32\pi}\int_0^1{dk^2}\int_0^\pi d\theta\sin^2\theta
\biggl[
2\sum_{ij}\frac{T^{sing}_{L,ij}(k^2,\cos^2\theta)-T^{quadr}_{L,ij}(k^2,\cos^2\theta)}{k^4}
%\eeq
%\[
\\
&+\sum_{ij}\frac{T^{sing}_{C,ij}(k^2,\cos^2\theta)
-T^{quadr}_{C,ij}(k^2,\cos^2\theta)}
{k^4}\biggr]
%\]
%\[
\\
&+\frac{3}{32\pi}\int_1^\infty{dk^2}\int_0^\pi d\theta\sin^2\theta
\biggl[
2\sum_{ij}\frac{T^{sing}_{L,ij}(k^2,\cos^2\theta)}{k^4}
+\sum_{ij}\frac{T^{sing}_{C,ij}(k^2,\cos^2\theta)}
{k^4}\biggr]
%\]
%\[
\\
&\equiv J^{sing<}+J^{sing>}=2\sum_{ij}J^<_{L,ij}+\sum_{ij}J^<_{C,ij}
+2\sum_{ij}J^>_{L,ij}+\sum_{ij}J^>_{C,ij},
%\]
\end{split}
\eeq

\noindent
where we have chosen $k=1$ to separate the regions of large and small momenta.

\begin{table}[tbhp!]
\caption{\label{isingle} Integrals $J^<$}
\begin{ruledtabular}
\begin{tabular}{|l|d||l|d|}
$J^{<}_{L,11}$ &0.\,042\,322\,(5) & $J^{<}_{C,11}$ & 0.\,174\,420\,(5)
\\
$J^{<}_{L,13}$ &
-\,0.\,008\,528\,(1)    & $J^{<}_{C,13}$ & -\,0.\,034\,331\,(5)
\\
$J^{<}_{L,21}$ & -\,0.\,251\,765\,(5)  & $J^{<}_{C,21}$ & 0.\,284\,129\,(8)
\\
$J^{<}_{L,22}$ & 0.\,020\,742\,(5)   & $J^{<}_{C,22}$ & -\,0.\,060\,654\,(3)
\\
$J^{<}_{L,41}$ & 0.\,050\,982\,(3) &       &
\\
$J^{<}_{L,42}$ & -\,0.\,002\,177\,(1) &&
\\
$J^{<}_{L,51}$ & 0.\,068\,860\,(3)    &&
\\
$J^{<}_{L,52}$ & -\,0.\,014\,977\,(1)   &&
\end{tabular}
\end{ruledtabular}
\end{table}

\begin{table}[tbhp!]
\caption{\label{singmore} Integrals $J^>$}
\begin{ruledtabular}
\begin{tabular}{|c|d||l|d|}
$J^{>}_{L,11}$ &
-\,0.\,728\,794\,(1)    & $J^{>}_{C,11}$ & -\,2.\,883\,062\,(2)
\\
$J^{>}_{L,13}$ & ~\,0.\,103\,289\,(1)    & $J^{>}_{C,13}$ &
0.\,412\,240\,(2)
\\
$J^{>}_{L,21}$ &
4.\,394\,262\,(2)      & $J^{>}_{C,21}$ & -\,1.\,142\,507\,(8)
\\
$J^{>}_{L,22}$ & -\,0.\,672\,947\,(5)    & $J^{>}_{C,22}$ &
0.\,205\,144\,(8)
\\
$J^{>}_{L,41}$ & -\,0.\,620\,559\,(3)
&&
\\
$J^{>}_{L,42}$ & 0.\,033\,714\,(7)  & &
\\
$J^{>}_{L,51}$ & -\,0.\,291\,783\,(4)    &&
\\
$J^{>}_{L,52}$ &
0.\,051\,666\,(6)       & &
\end{tabular}
\end{ruledtabular}
\end{table}

We collected the results for the individual integrals in \eq{singiniots} in Tables \ref{isingle} and \ref{singmore}. Summing the results in these tables we obtain

\beq
J^{sing<}=0.174\,48\,(1),          \qquad
J^{sing>}=1.129\,51\,(4).
\eeq

\noindent
We have checked by direct calculations that the sum

\beq \label{jsing}
J^{sing}=J^{sing<}+J^{sing>}=1.303\,99\,(5).
\eeq

\noindent
does not depend on the arbitrary separation point.

\subsection{Total Muon and Tauon Contributions}

Collecting the results in \eq{jreg} and \eq{jsing} we obtain

\beq
J=-0.842\,39\,(6),
\eeq

\noindent
and finally

\beq \label{muonloopcon}
\Delta E=-0.842\,39\,(6)\frac{\alpha^2(Z\alpha)}{\pi^3}\frac{m}{M}\widetilde E_F
\approx-0.2274~\mbox{Hz}.
\eeq

Using the same methods as above we also calculated a tiny contribution to hyperfine splitting generated by the tauon LBL scattering block in Fig. \ref{lblrec}

\beq \label{tauonloopcon}
\Delta E_\tau=-0.003\,58\,(1)\frac{\alpha^2(Z\alpha)}{\pi^3}\frac{m}{M}\widetilde E_F
\approx-0.0010~\mbox{Hz}.
\eeq

\section{Anomalous Magnetic Moments and Three-loop Contributions to Hyperfine Splitting}

Our final goal is to collect all three-loop radiative recoil corrections generated by the diagrams with closed fermion loops, but before that we would like to discuss calculation of the such corrections connected with the anomalous magnetic moments (AMM). Results for these three-loop radiative-recoil corrections  were reported in \cite{egs2005} but their derivation was never presented.

In our notation the classical Fermi result \cite{fermi1930} for the triplet-singlet splitting has the form

\beq \label{classic}
\Delta E=(1+a_e)(1+a_\mu)\widetilde E_F.
\eeq

\noindent
where $a_{e(\mu)}=(g_{e(\mu)}-2)/2$ are the electron and muon anomalous magnetic moments, respectively.

We see from \eq{classic} that the anomalous magnetic moments play a special role in the problem of hyperfine splitting and their contributions in certain cases  can be calculated exactly without expansion in small parameters. To avoid double counting we need to keep this fact in mind performing perturbative calculations. Physically it is more or less obvious that in the external field approximation only the total muon magnetic  moment matters. Our goal below is to clarify how this happens and to calculate the three-loop radiative-recoil corrections generated by the electron and muon AMMs. In some cases we will obtain results that are exact with respect to the lepton AMMs.

\subsection{AMM and Recoil and Radiative-Recoil Corrections}

The leading recoil correction to hyperfine splitting in muonium is generated by the diagrams  with two exchanged photons  in Fig.~\ref{twoph} and was calculated long time ago \cite{arn53,fm54,ns55}. The characteristic loop momenta in these diagrams are much larger than the electron mass, and therefore the leading recoil correction to hyperfine splitting can be calculated in the scattering approximation, ignoring the wave function momenta of order $mZ\alpha$ (see, e.g., \cite{egs2001,egs2007}). Let us recall the main steps in calculation of this leading recoil correction. In the scattering approximation the sum of the diagrams in Fig.~\ref{twoph} can be written as

\beq
-\frac{3}{8}\frac{(Z\alpha)mM}{\pi}\widetilde E_F
\int \frac{d^4 k}{i \pi^2k^4}\left[ L_{\mu\nu}^{(e)}(k) +
L_{\nu\mu}^{(e)}(-k) \right]  L_{\mu\nu}^{(\mu)}(-k),
\eeq

\noindent
where the electron skeleton factor $L_{\mu\nu}^{(e)}(k)$  is

\beq
L_{\mu\nu}^{(e)}(k)=-\frac{k^2}{k^4-4m^2k^2_0}\gamma^{\mu}\hat k \gamma^{\nu},
\eeq

\noindent
and the muon skeleton factor $L_{\mu\nu}^{(\mu)}(k)$ is obtained from the electron one by the substitution $m\to M$. Projecting the product of the fermion factors on hyperfine splitting\footnote{See an explicit expression for the projector in \cite{egs2004}, where there is a misprint in the overall sign of the projector.}  we obtain after the Wick rotation and transition to the four-dimensional spherical coordinates

\beq \label{angintsk}
\Delta E
=4\frac{(Z\alpha)mM}{\pi}\widetilde E_F
\int_0^\infty \frac{dk^2}{\pi}\int_0^\pi
d\theta\frac{\sin^2\theta(2+\cos^2\theta)}{(k^2+4m^2\cos^2\theta)(k^2+4M^2\cos^2\theta)}.
\eeq

\begin{figure}[htb]
\includegraphics
[height=1.5cm]
{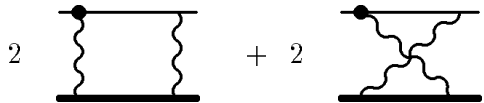}
\caption{\label{ammskel}
Skeleton diagrams with AMM insertions}
\end{figure}

Below we will need also the diagrams  in Fig.~\ref{ammskel} where one of the vertices in the skeleton diagrams in Fig.~\ref{twoph} is substituted by the  AMM. This substitution reduces to $\gamma^\mu\to -a_e\sigma^{\mu\lambda}k_\lambda/(2m)
=-\alpha/(2\pi)\sigma^{\mu\lambda}k_\lambda/(2m)$, where $k_\lambda$ is the momentum of the outgoing photon, see \cite{egs2004}. In the scattering approximation the sum of these diagrams has the form

\beq \label{ammuncm}
-\frac{3}{8}\frac{(Z\alpha)mM}{\pi}\widetilde E_F~
\int \frac{d^4 k}{i \pi^2k^4}\biggl[ L_{\mu\nu}^{(e,AMM)}(k) +
L_{\nu\mu}^{(e,AMM)}(-k) \biggr]  L_{\mu\nu}^{(\mu)}(-k),
\eeq

\noindent
where \cite{egs2004}\footnote{There is a misprint in the sign of this term  in \cite{egs2004}.}

\beq
L_{\mu\nu}^{(e,AMM)}(k) +
L_{\nu\mu}^{(e,AMM)}(-k)=-\frac{\alpha}{2\pi}
\frac{2k^2}{k^4-4m^2k^2_0}\left[\gamma^\mu\hat k \gamma^\nu
-k_0\left(\gamma^\mu\gamma^\nu-\frac{k^\mu\hat k\gamma^\nu+\gamma^\mu\hat k k^\nu}{k^2}\right)\right].
\eeq

\begin{figure}[htb]
\includegraphics
[height=1cm]
{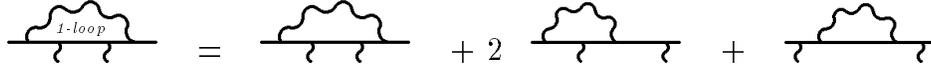}
\caption{\label{ff}
One-loop fermion factor}
\end{figure}

We need to clarify at this point why we consider the contribution of the AMM separately from the total one-loop electron factor in Fig.~\ref{ff}. The total electron factor can be represented as a sum

\beq \label{sumamm}
L_{\mu\nu}^{(e)}(k)=L_{\mu\nu}^{(e,AMM)}(k)+L_{\mu\nu}^{(e,r)}(k),
\eeq

\noindent
where $L_{\mu\nu}^{(e,r)}(k)$ is just the difference between the total one-loop electron factor $L_{\mu\nu}^{(e)}(k)$ and  $L_{\mu\nu}^{(e,AMM)}(k)$. First notice that the separation in \eq{sumamm} is gauge invariant since both the total one-loop electron factor and the AMM electron factor $L_{\mu\nu}^{(e,AMM)}(k)$ are gauge invariant. Different low momentum behavior of the two terms on the RHS in \eq{sumamm}  makes separate consideration of these terms convenient and even necessary from the calculational point of view. Due to the generalized low-energy theorem \cite{l1954,gmg1954,egs2001,egs2007} all terms linear in the small momentum $k$ are connected only with the term $L_{\mu\nu}^{(e,AMM)}(k)$, while the term $L_{\mu\nu}^{(e,r)}(k)$ decreases at least as $k^2$ at small $k^2$. This different low-energy behavior determines the structure of the integrals for the contributions to hyperfine
splitting and in many cases leads to qualitative differences between the contributions to HFS generated by the factors $L_{\mu\nu}^{(e,AMM)}(k)$ and $L_{\mu\nu}^{(e,r)}(k)$.

Projecting the expression in \eq{ammuncm} on HFS we obtain after the Wick rotation

\beq
\Delta E_{AMM}=\frac{\alpha(Z\alpha)mM}{\pi^2}\widetilde E_F
\int \frac{d^4 k}{
\pi^2}\frac{(2k^2+k_0^2)+3k_0^2}{(k^4+4m^2k_0^2)(k^4+4M^2k_0^2)},
\eeq

\noindent
or in four-dimensional spherical coordinates

\beq \label{angintamm}
\Delta E_{AMM}
=\frac{\alpha(Z\alpha)}{\pi^2}\widetilde E_F
(2mM)\int_0^\infty \frac{dk^2}{
\pi}\int_0^\pi
d\theta\frac{\sin^2\theta(2+4\cos^2\theta)}
{(k^2+4m^2\cos^2\theta)(k^2+4M^2\cos^2\theta)}.
\eeq

We see that the integrals in \eq{angintsk} and \eq{angintamm} can be calculated in terms of an auxiliary integral

\beq \label{auilintone}
\begin{split}
4mM\int_0^\infty \frac{dk^2}{
\pi}\int_0^\pi
d\theta\frac{\sin^2\theta(2+\xi\cos^2\theta)}{(k^2+4m^2\cos^2\theta)
(k^2+4M^2\cos^2\theta)}
%\eeq
%\[
\\
=\frac{ mM}{M^2-m^2} \int_0^\infty
\frac{dk^2}{k^2}
\left[F(\mu k,\xi) - F\left(\frac{k}{2}, \xi\right)\right],
\end{split}
\eeq
%\]

\noindent
where

\begin{equation} \label{AMM-PV-2}
F(\mu k,\xi) = \frac{2}{\mu k}\left(\sqrt{1+\mu^2k^2}-\mu k\right)
+ \xi\left(-\mu k\sqrt{1+\mu^2k^2}+\mu^2 k^2 +\frac12\right),
\end{equation}

\noindent
and we rescaled the integration variable $k\to k m$ so that it is  dimensionless on the right hand side in \eq{auilintone}.

Then the integrals in \eq{angintsk} and \eq{angintamm} can be written as

\beq \label{skelsing}
\Delta E
=\frac{Z\alpha}{\pi}\widetilde E_F\frac{ mM}{M^2-m^2} \int_0^\infty
\frac{dk^2}{k^2}
\left[F(\mu k,\xi) -F\left(\frac{k}{2} ,\xi\right)\right]_{|\xi=1},
\eeq

\noindent
and

\beq \label{ammdiv}
\Delta E_{AMM}
=\frac{\alpha(Z\alpha)}{2\pi^2}\widetilde E_F\frac{ mM}{M^2-m^2} \int_0^\infty
\frac{dk^2}{k^2}
\left[F(\mu k ,\xi) -F\left(\frac{k}{2} ,\xi\right)\right]_{|\xi=4}.
\eeq

\noindent
Both these integrals are linearly infrared divergent due to the singular behavior of the function $F(k,\xi)$ at small $k$, $F(k,\xi)_{|k\to0}\to2/k$. In the case of $\Delta E$ this divergence indicates that the integral in \eq{skelsing} contains a contribution of the previous order in $Z\alpha$, namely the leading nonrecoil contribution $\widetilde E_F$. The integral in \eq{ammdiv} contains a similar infrared divergence that again corresponds to the contribution of the previous order in $Z\alpha$, and starts to build the contribution proportional to the anomalous magnetic moment in \eq{classic}.

Subtracting the linear divergence  ($F(\mu k ,\xi)\to \widetilde F(\mu k,\xi)=F(\mu k,\xi)-2/(\mu k)$ and $F({k}/{2},\xi)\to\widetilde F({k}/{2},\xi) =F({k}/{2},\xi)-4/k$) we easily calculate the finite integral

\beq
\int_0^{\infty} {\frac{dk^2}{k^2}}
\left[\widetilde F(\mu k ,\xi) -\widetilde F\left(\frac{k}{2}, \xi\right)\right]=(\xi-4)\ln\frac{M}{m}.
\eeq

\noindent
Substituting this integral in \eq{skelsing} and \eq{ammdiv} we obtain the leading recoil correction \cite{arn53,fm54,ns55}

\beq
\Delta E_{skel,rec}
=-\frac{Z\alpha}{\pi}\widetilde E_F\frac{3 mM}{M^2-m^2}\ln\frac{M}{m},
\eeq

\noindent
and the recoil contribution of AMM

\beq
\Delta E_{AMM,rec}=0.
\eeq

\noindent
This nullification of the radiative-recoil contribution of AMM was discovered in \cite{sty1983}. We have obtained this result including only the one-loop contribution to AMM but it is easy to see that it holds even for an exact AMM, since higher order corrections to AMM change only the coefficient before the AMM vertex $-a_e\sigma^{\mu\lambda}k_\lambda/(2m)$.

\begin{figure}[htb]
\centering
\includegraphics[height=1.5cm]{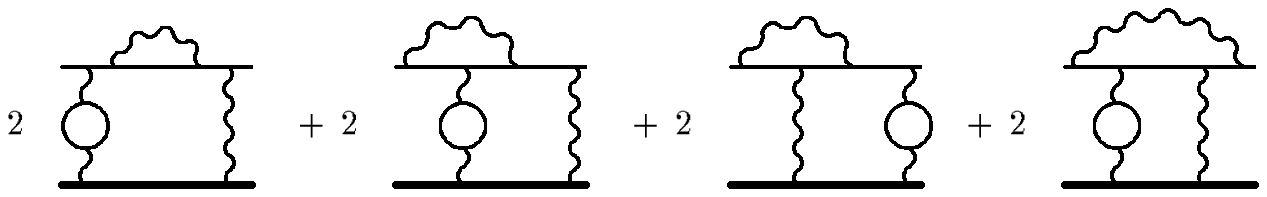}
\caption{Electron Polarization and Electron AMM}
\label{a}

\vspace{0.5cm}
\includegraphics[height=1.5cm]{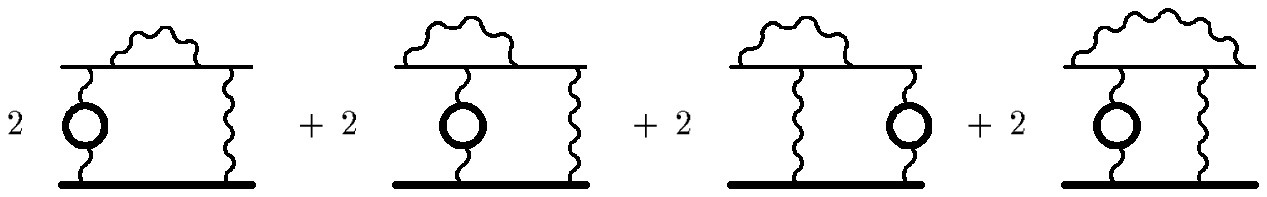}
\caption{Muon Polarization and Electron AMM}
\label{b}

\vspace{0.6cm}
\includegraphics[height=1.5cm]{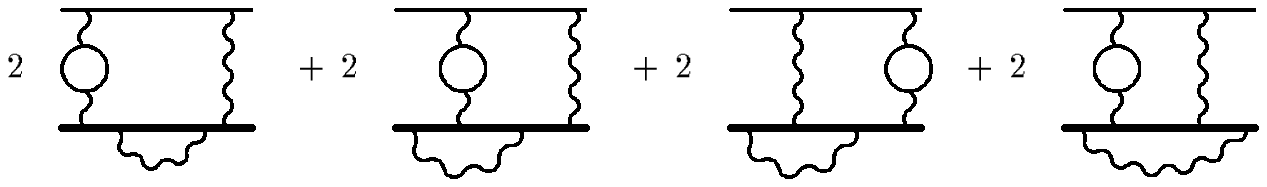}
\caption{Electron Polarization and Muon AMM}
\label{c}

\vspace{0.5cm}
\includegraphics[height=1.5cm]{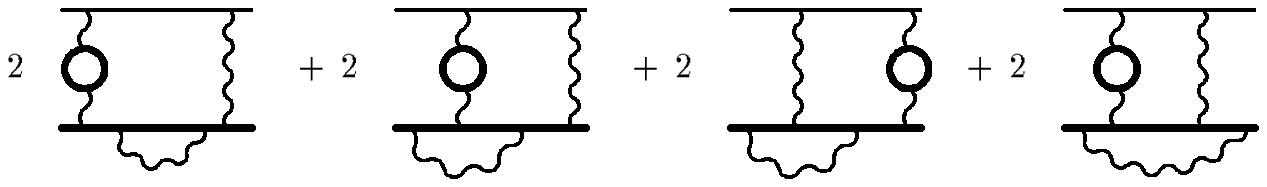}
%\vspace{-0.2cm}
\caption{Muon Polarization and Muon AMM}
\label{d}
\end{figure}

Let us turn now to the principal subject of our interest, contributions to HFS generated by the diagrams in Figs.~\ref{a}-\ref{d} with simultaneous insertions of fermion factors and electron or muon polarizations in the two-photon exchange diagrams in Fig.~\ref{twoph}\footnote{Below we omit the diagrams with the crossed photon lines in the figures.}. We consider below only the AMM contributions generated by these graphs. The contributions generated by the soft parts of the fermion factors $L_{\mu\nu}^{(e)}(k)$ and  $L_{\mu\nu}^{(e,AMM)}(k)$ were calculated in \cite{egs2003}.

\begin{figure}[htb]
\includegraphics
[height=1.5cm]
{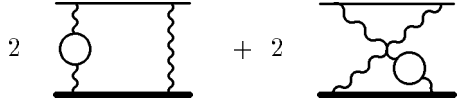}
\caption{\label{polarsk}
Diagrams with electron polarization insertions in the skeleton graphs}
\end{figure}

For normalization and illustrative purposes we will consider the contribution to HFS of the diagrams in Figs.~\ref{a}-\ref{d} in parallel with the diagrams with only the polarization insertions in Fig.~\ref{polarsk}, without radiation insertions in the fermion lines. The contribution to HFS of the diagrams in Fig.~\ref{polarsk} with the electron polarization insertions is obtained from the expression in \eq{skelsing} by insertion of the electron polarization operator $(\alpha/\pi)k^2I_{1e}(k^2)$, where (recall that the dimensionless momentum $k$ is measured in electron masses)

\beq
I_{1e}(k^2)=\int_0^1dv\frac{v^2\left(1-\frac{v^2}{3}\right)}{4+k^2(1-v^2)}.
\eeq

\noindent
Multiplying also by the combinatorial factor 2 we obtain

\beq \label{eepol}
\begin{split}
\Delta E_{epol}
&=\frac{\alpha(Z\alpha)}{\pi^2}\widetilde E_F\frac{2mM}{M^2-m^2}
\int_0^{\infty} \frac{dk^2}{k^2}
\left[F(\mu k ,\xi) -F\left(\frac{k}{2},\xi\right)\right]_{|\xi=1}k^2I_{1e}(k^2)
%\eeq
%\[
\\
&=\frac{\alpha(Z\alpha)}{\pi^2}\widetilde E_FJ_{e}(\xi)_{|\xi=1},
\end{split}
\eeq
%\]

\noindent
where we have introduced an auxiliary integral

\begin{equation} \label{auxil}
J_{e}(\xi)=\frac{2mM}{M^2-m^2}\int_0^{\infty} \frac{dk^2}{k^2}
\left[F(\mu k ,\xi) -F\left(\frac{k}{2} ,\xi\right)\right]k^2I_{1e}(k^2).
\end{equation}

Similarly the contribution to HFS generated by the diagrams in  Fig.~\ref{a} with AMM and electron vacuum polarizations can be obtained by the insertion of the polarization operator in the the integral in \eq{ammdiv} (an extra factor 2 is again due to combinatorics)

\beq \label{epolamm}
\begin{split}
\Delta E^a_{AMM}
&=\frac{\alpha^2(Z\alpha)}{\pi^3}\widetilde E_F\frac{mM}{M^2-m^2} \int_0^\infty
\frac{dk^2}{k^2}\left[F(\mu k,\xi)-F\left(\frac{k}{2}, \xi\right)\right]_{|\xi=4}k^2I_{1e}(k^2)
%\eeq
%\[
\\
&=\frac{\alpha^2(Z\alpha)}{\pi^3}\widetilde E_F\left(\frac{1}{2}J_{e}(\xi)_{|\xi=4}\right).
\end{split}
\eeq
%\]

We see that both the contributions of the diagrams in Fig.~\ref{polarsk} and in Fig.~\ref{a} (as well as with some modifications of the other diagrams in  Figs.~\ref{b}-\ref{d}) can be calculated in terms of the integral $J_{e}(\xi)$. Let us calculate this auxiliary integral. We represent it as a sum of terms

\beq
J_e(\xi)=J_e(\xi)|_{NR}+J_e(\xi)|_{recoil}
=J_e(\xi)|_{NR}+J'_e(\xi)|_{recoil}+J''_e(\xi)|_{recoil},
\eeq

\noindent
where $J_e(\xi)|_{NR}$ is the integral of the leading term when $\mu\to0$, namely the term $2/(\mu k)$ in $F(\mu k,\, \xi)$, $J'_e(\xi)|_{recoil}$ is the integral of $\widetilde F(\mu k,\, \xi)=F(\mu k,\, \xi)-2/(\mu k)$, and $J''_e(\xi)|_{recoil}$ is the integral of $[-F({k},{2}/\xi)]$.

It is easy to see that

\begin{equation} \label{jnonrec}
J_{e}|_{NR}
=\frac{16M^2}{M^2-m^2}\int_0^{\infty} dkI_{1e}(k^2)
=\frac{M^2}{M^2-m^2}\frac{3\pi^2}{4}.
\eeq

Next we calculate the integral

\beq
J'_e(\xi)|_{recoil}=
\frac{2mM}{M^2-m^2}\int_0^{\infty} {\frac{dk^2}{k^2}}
\widetilde F(\mu k,\xi) k^2I_{1e}(k^2),
\eeq

\noindent
with linear accuracy in the small parameter $\mu$. As usual with the integrals of this type (see, e.g., \cite{egs2001,egs2007,egs1998,blp}) we introduce an auxiliary parameter $\sigma$ that satisfies the inequality $1\ll \sigma\ll \mu^{-1}$. The parameter $\sigma$ is used to separate the momentum integration into two regions, a region of small momenta $0\leq k\leq\sigma$, and a region of large momenta $\sigma\leq k<\infty$. In the region of small momenta one uses the condition $\mu k\ll 1$ to simplify the integrand, and in the region of large momenta the same goal is achieved with the help of the condition $k\gg 1$. For $k\simeq \sigma$ both conditions on the integration
momentum are valid simultaneously, so in the sum of the low-momenta and
high-momenta integrals all $\sigma$-dependent terms cancel and one
obtains a $\sigma$-independent result for the total momentum integral. Calculation of the integral in the small momentum region $\mu k\ll 1$ is straightforward, and  we obtain

\beq \label{jeless}
\begin{split}
J'_{e}(\xi)|_{recoil}^{<}
&=\frac{2mM}{M^2-m^2}\int_0^{\sigma^2} \frac{dk^2}{k^2}
\widetilde F(\mu k ,\xi) k^2I_{1e}(k^2)
%\eeq
%\[
\\
&=\frac{mM}{M^2-m^2}\left[(\xi-4)\left(\frac{2}{3}\ln^2{\sigma} -\frac{10}{9}\ln{\sigma}
+\frac{28}{27}\right)+{\cal O}(\mu\sigma)\right].
\end{split}
\eeq
%\]

\noindent
Calculation of the contribution from large integration momenta

\beq
J'_{e}(\xi)|_{recoil}^{>}
=\frac{2mM}{M^2-m^2}\int_{\sigma^2}^\infty {\frac{dk^2}{k^2}}~
\widetilde F(\mu k,\xi) k^2I_{1e}(k^2)
\eeq

\noindent
is a bit more involved. In this region we use the well known leading terms in the asymptotic expansion of the polarization operator $k^2I_{1e}(k^2)_{|k\to\infty}\to (2/3)\ln k-5/9$, and represent the subtracted weight function $\bar{F}(\mu k,\xi)$ in the form

\begin{equation} \label{AMM-PV-11}
\begin{split}
\widetilde F(\mu k,\xi)& =\frac{2}{\mu k}\left(\sqrt{1+\mu^2k^2}-\mu k -1\right)
+\xi\left(-\mu k\sqrt{1+\mu^2k^2}+\mu^2 k^2+\frac{1}{2}\right)
%\end{equation}
%\[
\\
&=2\Phi^{\mu}_0(k) + \xi\Phi^{\mu}_1(k),
\end{split}
\eeq
%\]

\noindent
where the functions $\Phi^{\mu}_{0,1}(k)$  are defined in Appendix A of \cite{egs1998}. In these terms

\beq
\begin{split}
J'_{e}(\xi)|_{recoil}^>
&=\frac{2mM}{M^2-m^2}\left\{\int_{\sigma^2}^\infty \frac{dk^2}{k^2}
\left[2\Phi^{\mu}_0(k) + \xi\Phi^{\mu}_1(k)\right]
\left(\frac{2}{3}\ln{k}- \frac{5}{9}\right)+{\cal O}\left(\frac{1}{\sigma^2}\right)\right\}
%\eeq
%\[
\\
&=\frac{2mM}{M^2-m^2}\left[\frac{4}{3}V_{110}+ \xi\frac{2}{3}V_{111}
- \frac{10}{9}V_{100} - \xi\frac{5}{9}V_{101}\right],
\end{split}
\eeq
%\]

\noindent
where the integrals $V_{mnl}$ were defined in Appendix C of \cite{egs1998}.
These integrals were calculated in the limit of small $\mu\sigma$, and using these results we obtain

\beq \label{jemore}
\begin{split}
J'_{e}(\xi)|_{recoil}^>
=\frac{mM}{M^2-m^2}\biggl[&(\xi-4)\left(-\frac{2}{3}\ln^2\sigma
+\frac{10}{9}\ln\sigma+\frac{\pi^2}{9}+\frac{2}{3}\ln^2(2\mu)\right)
\\
&
+\frac{8(1+2\xi)}{9}\ln(2\mu)+\frac{8(\xi-1)}{9}\biggr].
\end{split}
\eeq

\noindent
In the sum of the contributions of small \eq{jeless} and large \eq{jemore} integration momenta regions all $\sigma$-dependent terms cancel and we obtain

\beq \label{jeprime}
J'_{e}(\xi)|_{recoil}=\frac{mM}{M^2-m^2} \left[\frac{2(\xi-4)}{3}\ln^2{\frac{M}{m}}
- \frac{8(1+2\xi)}{9}\ln{\frac{M}{m}}
+\frac{\pi^2(\xi-4)}{9} +\frac{4(13\xi-34)}{27}\right].
\eeq

\noindent
Calculation of the $\mu$-independent integral $J''_{e}(\xi)$ is straightforward and we obtain

\beq \label{jtwoprimes}
\begin{split}
J''_{e}(\xi)|_{recoil}& = -\frac{2mM}{M^2-m^2}\int_0^{\infty} \frac{dk^2}{k^2}F\left(\frac{k}{2},\xi\right)k^2I_{1e}(k^2)
%\eeq
%\[
\\
&=\frac{mM}{M^2-m^2}\left[\frac{(\xi-4)\pi^2}{9} - \frac{2(17\xi-32)}{27}\right].
\end{split}
\eeq
%\]

Collecting the results in \eq{jnonrec}, \eq{jeprime}, and \eq{jtwoprimes} we obtain the auxiliary integral $J_{e}(\xi)$ with linear accuracy in the small parameter $m/M$

\beq \label{jexi}
\begin{split}
J_{e}(\xi)
&=\frac{3\pi^2}{4}\frac{M^2}{M^2-m^2}
\\
&+
\frac{mM}{M^2-m^2}\biggl[\frac{2(\xi-4)}{3}\ln^2{\frac{M}{m}}
-\frac{8(1+2\xi)}{9}\ln{\frac{M}{m}}
+\frac{2\pi^2(\xi-4)}{9} +\frac{2(\xi-4)}{3}\biggr].
\end{split}
\eeq

We need to consider also the diagrams with insertions of the muon vacuum polarizations. In this case the integrals in \eq{eepol} and \eq{epolamm} turn into

\beq
\Delta E_{mupol}
=\frac{\alpha(Z\alpha)}{\pi^2}\widetilde E_F\frac{2mM}{M^2-m^2}
\int_0^{\infty} \frac{dk^2}{k^2}
\left[F(\mu k,\xi)-F\left(\frac{k}{2},\xi\right)\right]_{|\xi=1}k^2I_{1\mu}(k^2),
\eeq

\noindent
and

\beq
\Delta E^b_{AMM}
=\frac{\alpha^2(Z\alpha)}{\pi^3}\widetilde E_F\frac{mM}{M^2-m^2} \int_0^\infty
\frac{dk^2}{k^2}
\left[F(\mu k,\xi)-F\left(\frac{k}{2},\xi\right)\right]_{|\xi=4}k^2I_{1\mu}(k^2),
\eeq

\noindent
where (recall the the dimensionless momentum $k$ is measured in electron masses).

\beq
k^2I_{1\mu}(k^2)=k^2m^2\int_0^1 dv\frac{v^2\left(1-\frac{v^2}{3}\right)}{4M^2+k^2m^2(1-v^2)}.
\eeq

\noindent
Let us rescale the dimensionless integration momentum once more $k\to(M/m)k=k/(2\mu)$. After this rescaling

\beq
k^2I_{1\mu}(k^2)\to k^2\int_0^1dv\frac{v^2\left(1-\frac{v^2}{3}\right)}{4+k^2(1-v^2)}
=k^2I_{1e}(k^2),
\eeq

\beq
\begin{split}
\Delta E_{mupol}
&=\frac{\alpha(Z\alpha)}{\pi^2}\widetilde E_F\frac{2mM}{M^2-m^2}
\int_0^{\infty} \frac{dk^2}{k^2}
\left[F\left(\frac{k}{2},\xi\right)-F\left(\frac{k}{4\mu},\xi\right)
\right]_{|\xi=1}k^2I_{1e}(k^2)
%\eeq
%\[
\\
&=\frac{\alpha(Z\alpha)}{\pi^2}\widetilde E_FJ_{\mu}(\xi)_{|\xi=1},
\end{split}
\eeq
%\]

\noindent
and

\beq
\begin{split}
\Delta E^b_{AMM}
&=\frac{\alpha^2(Z\alpha)}{\pi^3}\widetilde E_F\frac{mM}{M^2-m^2} \int_0^\infty
\frac{dk^2}{k^2}
\left[F\left(\frac{ k}{2},\xi\right)-F\left(\frac{k}{4\mu},\xi\right)\right]_{|\xi=4}
k^2I_{1e}(k^2)
%\eeq
%\[
\\
&=\frac{\alpha^2(Z\alpha)}{\pi^3}\widetilde E_F\left(\frac{1}{2}J_{\mu}(\xi)_{|\xi=4}\right),
\end{split}
\eeq
%\]

\noindent
where

\beq
J_{\mu}(\xi)=\frac{2mM}{M^2-m^2}\int_0^{\infty} \frac{dk^2}{k^2}
\left[F\left(\frac{k}{2},\xi\right)-F\left(\frac{k}{4\mu},\xi\right)
\right] k^2I_{1e}(k^2).
\eeq

\noindent
The integral with the function $F({k}/{2},\xi)$ is exactly the integral with the same function that we calculated above in the electron case, but with an opposite sign, see \eq{jtwoprimes}. One can check that the integral with the function $F({k}/(4\mu),\xi)$ does not generate contributions linear in the mass ratio. Then with the linear in the mass ratio accuracy we obtain

\beq \label{jmuxi}
J_{\mu}(\xi)
=-\frac{mM}{M^2-m^2}
\left[\frac{(\xi-4)\pi^2}{9} - \frac{2(17\xi-32)}{27}\right].
\eeq

We are now ready to use the integrals in \eq{jexi} and \eq{jmuxi} for calculation of the radiative-recoil corrections. Let us consider first corrections due to one-loop polarization insertions in the exchanged photons in Fig.~\ref{polarsk}. Collecting the integrals in \eq{jexi} and \eq{jmuxi} we reproduce the well known result \cite{sty1983}

\beq
\Delta E_{pol}=\Delta E_{epol}+\Delta E_{mupol}
=\frac{\alpha(Z\alpha)}{\pi^2}\widetilde E_F\left\{
\frac{3\pi^2}{4}
+\frac{m}{M}\left[-2\ln^2\frac{M}{m}
-\frac{8}{3}\ln{\frac{M}{m}} -\frac{\pi^2}{3} - \frac{28}{9}\right]\right\}.
\eeq

Let us turn to the diagrams with AMM insertions in Figs.~\ref{a}-\ref{d}. Up to this moment we considered only calculations of the diagrams with AMM insertions in the electron line. But from the calculations above it is clear that AMM enters only as a common factor and the integrals are identical for insertions of the electron and muon AMMs. Hence, the contribution of Fig.~\ref{c} coincides with the contribution of Fig.~\ref{a} and the contribution of Fig.~\ref{d} coincides with the contribution of Fig.~\ref{b}. As a result the total contribution to HFS with insertions of both the electron and muon AMM is twice the result with only the electron  AMM.

Now we are ready to present the results for the contributions to HFS generated by the diagrams with AMM insertions in Figs.~\ref{a}-\ref{d}. We obtain

\beq \label{amma}
\Delta E^a_{AMM}=
\frac{\alpha^2(Z\alpha)}{\pi^3}\widetilde E_F\left(\frac{1}{2}J_{e}(\xi)_{|\xi=4}\right)
=\frac{\alpha^2(Z\alpha)}{\pi^3}\widetilde E_F
\left[\frac{3\pi^2}{8}+ \frac{m}{M}\left(-4\ln{\frac{M}{m}}\right)\right],
\eeq

\beq  \label{ammb}
\Delta E^b_{AMM}=
\frac{\alpha^2(Z\alpha)}{\pi^3}\widetilde E_F\left(\frac{1}{2}J_{\mu}(|\xi)_{\xi=4}\right)
=\frac{4}{3}\frac{\alpha^2(Z\alpha)}{\pi^3}\frac{m}{M}\widetilde E_F,
\eeq

\beq  \label{ammc}
\Delta E^c_{AMM}=
\frac{\alpha(Z^2\alpha)(Z\alpha)}{\pi^3}\widetilde E_F\left(\frac{1}{2}J_{e}(|\xi)_{\xi=4}\right)
=\frac{\alpha(Z^2\alpha)(Z\alpha)}{\pi^3}\widetilde E_F
\left[\frac{3\pi^2}{8}+\frac{m}{M}\left(-4\ln{\frac{M}{m}}\right)\right],
\eeq

\beq \label{ammd}
\Delta E^d_{AMM}=
\frac{\alpha(Z^2\alpha)(Z\alpha)}{\pi^3}\widetilde E_F\left(\frac{1}{2}J_{\mu}(\xi)_{|\xi=4}\right)
=\frac{4}{3}\frac{\alpha(Z^2\alpha)(Z\alpha)}{\pi^3}
\frac{m}{M}\widetilde E_F.
\eeq

\subsection{Discussion of Three-Loop AMM Contributions}

Usually, in discussions of the nonrecoil corrections to HFS, the muon AMM is included in the definition of the Fermi energy, $E_F=(1+a_\mu)\widetilde E_F$. As a result of this convention some of the contributions generated by the AMM in \eq{amma}-\eq{ammd} are already taken into account in the standard compilations  of all corrections to HFS. Our next task is to figure out what entries in \eq{amma} are new. The nonrecoil contribution in \eq{amma} was calculated long time ago, see reviews in \cite{egs2001,egs2007,mtn2012}. The nonrecoil contribution in \eq{ammc} is taken into account when one writes the classical nonrecoil Kroll-Pollack contribution  \cite{kp1952,kks1951,kk1952} in terms of the Fermi energy $E_F$, see, e.g., \cite{egs2001,egs2007,mtn2012}. The three-loop radiative-recoil terms in \eq{amma}-\eq{ammd} were obtained in \cite{egs2005} (see also \cite{es2009})

\beq \label{ammradrec}
\Delta E_{rec}=
\frac{\alpha^2(Z\alpha)}{\pi^3}\widetilde E_F
\frac{m}{M}\left(-4\ln{\frac{M}{m}}+\frac{4}{3}\right)
+\frac{\alpha(Z^2\alpha)(Z\alpha)}{\pi^3}\widetilde E_F
\frac{m}{M}\left(-4\ln{\frac{M}{m}}+\frac{4}{3}\right).
\eeq

They were not included in \cite{egs2003}, where only the contributions connected with the soft parts of the fermion factors $L_{\mu\nu}^{(e,r)}(k)$ and $L_{\mu\nu}^{(\mu,r)}(k)$ were considered. Only the one-loop anomalous magnetic moments of electron and muon are accounted for in \eq{ammradrec}. However, as is obvious from the derivation of this contribution, one can account for the AMMs in this expression exactly by the trivial substitutions $\alpha/(2\pi)\to a_e$ and $Z^2\alpha/(2\pi)\to a_\mu$ in the first and second terms on the RHS in \eq{ammradrec}.

\section{Summary}

The results presented above conclude calculation of all radiative-recoil corrections to HFS of order $\alpha^2(Z\alpha)(m/M)\widetilde E_F$ generated by the diagrams with closed fermion loops. The leading logarithm cubed and logarithm squared contributions are well known (see, e.g., reviews \cite{egs2001,egs2007,mtn2012}) and below we collect all hard single-logarithmic and nonlogarithmic three-loop radiative-recoil corrections that arise due to the diagrams with closed fermion loops. Consider first the single-logarithmic corrections. They were calculated in  \cite{egs2001rr,egs2003,egs2004,egs2005,jetp2010,prd2009pr,es2013}\footnote{The AMM contributions in \eq{ammradrec} were not included in the result in \cite{egs2003}. The logarithmic (and nonlogarithmic) results from \cite{egs2001rr,egs2003,egs2004} were collected in \cite{egs2005}, where they were amended by the AMM contributions in \eq{ammradrec}. Contributions of the subtracted electron (muon) factor and the respective AMM were written in \cite{egs2005} separately. After calculation of the new contributions in \cite{jetp2010,prd2009pr} a new collection of all known results was presented in \cite{es2009}. All expressions for logarithmic contributions in \cite{es2009} are correct. The results for the diagrams with the electron and muon factors were written in \cite{es2009} for the full electron and muon factors including the AMMs (sums of the respective contributions in \cite{egs2005}). Unfortunately, there were no reference to \cite{egs2005} in \cite{es2009} and there was only reference to \cite{egs2003}, where the AMMs were not included. So if a reader wanted to check the correct results in  \cite{es2009} by going to \cite{egs2003} he/she would discover a (fake) discrepancy between \cite{es2009} and \cite{egs2003}. Once again, the full results with AMMs are in \cite{egs2005} and \cite{es2009}.} ($Z=1$ below)

\beq \label{totlog}
\Delta E_{log}=
\frac{\alpha^3}{\pi^3}\frac{m}{M}\widetilde E_F\biggl(-6\pi^2\ln2+\frac{\pi^2}{3}+\frac{27}{8}\biggr)\ln{\frac{M}{m}}.
\eeq

\noindent
This is the total single-logarithmic contribution first calculated in \cite{es2013}.

Let us turn to the nonlogarithmic contributions, including the muon loop result in \eq{muonloopcon}. They were calculated in \cite{egs2001rr,egs2003,egs2004,egs2005,jetp2010,prd2009pr,es2014,
es2014_2}\footnote{Let us mention that the term with the fourth power of logarithm  from \cite{egs2001rr} was swallowed in the nonlogarithmic "constant" in \cite{es2009}. Here we choose to write it separately. Also notice that the nonlogarithmic result in \cite{egs2004} was later corrected, see Erratum in \cite{egs2004}. In addition we have slightly improved the contribution from \cite{prd2009pr}. Considering eq.(26) and (25) from \cite{prd2009pr} we see that the result in eq.(27) in \cite{prd2009pr} can be written with an extra digit (but we do not change
the error bars). Then the constant in eq.(32) in \cite{prd2009pr} that is a sum of the improved number in eq.(27) in \cite{prd2009pr} and the number in eq.(29) in \cite{prd2009pr} is 11.2958(20). We use this number with its original error bars as the result of \cite{prd2009pr}.}

\beq \label{nontlog}
\Delta E_{nonlog}=68.507~(2)\frac{\alpha^3}{\pi^3}\frac{m}{M}\widetilde E_F
\eeq

Then the sum of all single-logarithmic and nonlogarithmic three-loop radiative-recoil corrections \cite{es2009,es2013,es2014,es2014_2}\footnote{We do not include the term with the fourth power of logarithm from \cite{egs2001rr} and  corrections due to the tauon and hadron loop light-by-light scattering blocks in the expression below, see \cite{es2014_2}.} has the form

%\begin{eqnarray}
\beq \label{singlnon}
\Delta E_{tot}=\frac{\alpha^3}{\pi^3}\frac{m}{M}\widetilde E_F
%&&
\biggl[\biggl(-6\pi^2\ln2+\frac{\pi^2}{3}+\frac{27}{8}\biggr) \ln{\frac{M}{m}}
%\nonumber\\*
%\eeq
%\[
%&&
+68.507~(2)\biggr].
%\label{finres}
\eeq
%\end{eqnarray}

\noindent
Numerically this contribution to HFS in muonium is

\begin{equation}
\Delta E_{tot}=-30.99~\mbox{Hz}.
\end{equation}

For completeness let us cite also some other minor radiative-recoil corrections. They are the leading large logarithm quadrupled contribution \cite{egs2001rr}

\beq
\Delta E=-\frac{8}{9}\frac{\alpha^4}{\pi^4}\frac{m}{M}\widetilde E_F\ln^4\frac{M}{m}
=-0.4504~{\mbox Hz},
\eeq

\noindent
the tauon light-by-light contribution in \eq{tauonloopcon}

\beq
\Delta E_\tau=-0.003~58(1)\frac{\alpha^3}{\pi^3}\frac{m}{M}\widetilde E_F
=-0.0010~{\mbox Hz},
\eeq

\noindent
and the hadron light-by-light contribution \cite{ksv2008}

\beq
\Delta E=-0.0065~{\mbox Hz}.
\eeq

\noindent
The result in \eq{singlnon} includes all already known hard three-loop single-logarithmic and nonlogarithmic corrections of order $\alpha^2(Z\alpha)(m/M)\widetilde E_F$ to HFS in muonium. There are still two gauge invariant sets of diagrams that generate such corrections and remain uncalculated. These are the diagrams with two radiative photon insertions in one and the same fermion line, either electron or muon. Using the known results for the respective diagrams with one radiative photon insertion in either of the fermion lines (see, e.g., reviews \cite{egs2001,egs2007}) and the result in \cite{egs2004} for the diagrams with simultaneous insertions of radiative photons in both fermion lines we estimate the contribution of these diagrams to be about 10-15 Hz. Calculation of these diagrams is the next task for the theory.

\acknowledgments

This work was supported by the NSF grants PHY-1066054 and PHY-1402593. The work of V. S. was supported in part by the RFBR grant 14-02-00467 and by the DFG Grant
No. HA 1457/9-1.


\begin{thebibliography}{99}

\bibitem{egs2001} M.~I.~Eides, H.~Grotch, and V.~A.~Shelyuto, Phys. Rep.
\textbf{342}, 63 (2001).

\bibitem{egs2007} M.~I.~Eides, H.~Grotch, and V.~A.~Shelyuto, \textit{Theory
of Light Hydrogenic Bound States}, (Springer, Berlin, Heidelberg, New York, 2007).

\bibitem{mtn2012} P.~J.~Mohr, B.~N.~Taylor, and D.~B.~Newell, Rev. Mod. Phys. \textbf{84}, 1527 (2012).

\bibitem{mbb} F.~G.~Mariam, W.~Beer, P.~R.~Bolton et al, Phys. Rev.
Lett. \textbf{49}, 993 (1982).

\bibitem{lbdd} W.~Liu, M.~G.~Boshier, S.~Dhawan et al, Phys. Rev. Lett. \textbf{82}, 711 (1999).

\bibitem{sasaki} K.~Sasaki et al, J. of Physics: Conference Series \textbf{408}, 012074 (2013).

\bibitem{tanaka} K.~S.~Tanaka et al, JPS Conf. Proc. \textbf{2}, 010405 (2014).

\bibitem{egs2001rr} M.~I.~Eides, H.~Grotch, and V.~A.~Shelyuto, Phys. Rev. D
\textbf{65}, 013003 (2001).

\bibitem{egs2003} M.~I.~Eides, H.~Grotch, and V.~A.~Shelyuto, Phys. Rev. D
\textbf{67}, 113003 (2003).

\bibitem{egs2004} M.~I.~Eides, H.~Grotch, and V.~A.~Shelyuto, Phys. Rev. D
\textbf{70}, 073005 (2004); Erratum, Phys. Rev. D \textbf{88}, 119901(E) (2013).

\bibitem{egs2005} M.~I.~Eides, H.~Grotch, and V.~A.~Shelyuto, Can. J. Phys. \textbf{83}, 363 (2005).

\bibitem{es2009} M.~I.~Eides and V.~A.~Shelyuto, Phys. Rev. Lett.
\textbf{103}, 133003 (2009).

\bibitem{prd2009pr} M.~I.~Eides and V.~A.~Shelyuto, Phys. Rev. D
\textbf{80}, 053008 (2009).

\bibitem{jetp2010} M.~I.~Eides and V.~A.~Shelyuto, JETP \textbf{110}, 17 (2010).

\bibitem{es2013} M.~I.~Eides and V.~A.~Shelyuto, Phys. Rev. D
\textbf{87}, 013005 (2013).

\bibitem{es2014} M.~I.~Eides and V.~A.~Shelyuto, Phys. Rev. D
\textbf{89}, 014034 (2014).

\bibitem{es2014_2} M.~I.~Eides and V.~A.~Shelyuto, Phys. Rev. Lett. \textbf{112}, 173004 (2014).

\bibitem{fermi1930} E.~Fermi, Z. Phys. \textbf{60}, 320 (1930).

\bibitem{arn53} R.~Arnowitt, Phys. Rev. \textbf{92}, 1002 (1953).

\bibitem{fm54}  T.~Fulton and P.~C.~Martin, Phys. Rev. \textbf{95}, 811 (1954).

\bibitem{ns55} W.~A.~Newcomb and E.~E.~Salpeter, Phys. Rev. \textbf{97}, 1146 (1955).

\bibitem{l1954} F.~E.~Low, Phys. Rev. \textbf{96}, 1428 (1954).

\bibitem{gmg1954} M.~Gell-Mann and M.~L.~Goldberger, Phys. Rev. \textbf{96}, 1433 (1954).

\bibitem{sty1983} J.~R.~Sapirstein, E.~A.~Terray, and D.~R.~Yennie, Phys. Rev. Lett. \textbf{51}, 982 (1983); Phys. Rev. D \textbf{29}, 2290 (1984).

\bibitem{egs1998} M.~I.~Eides, H.~Grotch, and V.~A.~Shelyuto, Phys. Rev. D
\textbf{58}, 013008 (1998).

\bibitem{blp} V.~B.~Berestetskii, E.~M.~Lifshitz, and L.~P.~Pitaevskii,
Quantum electrodynamics, 2nd Edition, Pergamon Press, Oxford, 1982.

\bibitem{kp1952} N.~Kroll and F.~Pollock, Phys. Rev. \textbf{84}, 594 (1951); ibid. \textbf{86}, 876 (1952).

\bibitem{kks1951} R.~Karplus, A.~Klein, and J.~Schwinger, Phys. Rev. \textbf{84}, 597 (1951).

\bibitem{kk1952} R.~Karplus and A.~Klein, Phys. Rev. \textbf{85}, 972 (1952).

\bibitem{ksv2008} S.~G.~Karshenboim, V.~A.~Shelyuto, and A.~I.~Vainshtein,
Phys. Rev. D \textbf{78}, 065036 (2008).











\end{thebibliography}
\end{document}